\documentstyle[12pt]{article}
\renewcommand\theequation{\thesection.\arabic{equation}}
\begin{document}
\setlength{\unitlength}{1mm}
\newcommand{\te}{\theta}
\newcommand{\bee}{\begin{equation}}
\newcommand{\ene}{\end{equation}}
\newcommand{\tra}{\triangle\theta_}

{\hfill Preprint JINR E2-93-46 } \vspace*{2cm} \\
\begin{center}
{\Large\bf Finite-Temperature Scalar Field Theory }\\
\end{center}
\begin{center}
{\Large\bf in Static de Sitter Space}
\end{center}

\bigskip\bigskip

\begin{center}
{\bf D.V. Fursaev}
\footnote{e-mail: fursaev@theor.jinrc.dubna.su}
\end{center}

\begin{center}
{\it Laboratory of Theoretical Physics, Joint Institute for\\
Nuclear Research, Head Post Office, P.O.Box 79, Moscow,\\
Russia}
\end{center}

\bigskip

\begin{center}
{\bf G. Miele}
\footnote{e-mail: miele@napoli.infn.it; miele@vaxna1.cern.ch}
\end{center}

\begin{center}
{\it Dipartimento di Scienze Fisiche, Universit\`a di Napoli - Federico II -,
and INFN Sezione di Napoli, Mostra D'Oltremare Pad. 19, 80125, Napoli,\\
Italy}
\end{center}

\bigskip\bigskip\bigskip

\begin{abstract}
The  finite-temperature one-loop effective potential for a scalar
field in the static de Sitter space-time is obtained.
Within this framework, by using zeta-function regularization, one can get,
in the
conformally invariant case, the explicit expression for the stress tensor
anomaly. Its value turns out to depend on the thermal state of the system.
This conclusion is different from the one
derived by other authors, who considered thermal properties of ultraviolet
divergences in static spaces
ignoring the effects of horizons. The behaviour
of the effective potential in the ground state and in de Sitter-invariant state
is also studied, showing the role played by the curvature on the minima.
\end{abstract}

\vspace*{2cm}

\begin{center}
{\it PACS number(s): 98.80Cq, 03.70+k}
\end{center}

\newpage
\baselineskip=.8cm

\section{Introduction}

The hypothesis that the early universe might have undergone an exponential
expansion might explain a number of essential questions. Why, for
example, the observed space is homogeneous and isotropic and energy density
in it is so close to the critical value \cite{a15}. In the
exponentially expanding epoch the
universe has the de Sitter geometry
with fixed radius. If the radius is sufficiently small, there may be
interesting effects arising from the behavior of quantum field theories
in such curved space. In this way gravitation can influence the
properties of the effective potential and can change the symmetry-breaking
pattern in gauge models.

In the one-loop approximation and assuming a de Sitter space-time
this problem has been studied for scalar electrodynamics \cite{a1}
and for the more realistic $SU(5)$ gauge theory
\cite{a14},\cite{a2}. These papers show that
gravitational effects change the phase structure of the theory, but
analysis there was restricted to a particular choice for the quantum
state of the system,i.e.
to the state which is invariant under transformations of the de Sitter group
\cite{a3}. All observers moving freely register it equally as a thermal
equilibrium state at the same temperature $(2\pi a)^{-1}$ (Hawking
temperature),
with $a$ the radius of the space \cite{a5}.

Note that the thermal equilibrium state in de Sitter space-time is always
possible in static coordinates where the external gravitational field does
not depend on time.
Thus, a natural question arises: how does symmetry breaking occur in
the de Sitter universe if a given quantum field is in an arbitrary thermal
equilibrium state different from the invariant one? For this purpose,
the study of finite-temperature quantum field theory in a static
de Sitter space-time is necessary.

This subject is also interesting by itself.
Let us recall that a freely moving observer in this space
has an event horizon separating from the whole space-time the region
he can never see.
The presence of horizons can have interesting consequences.
It is known, for instance, that there is a close connection between event
horizons and thermodynamics \cite{a5}. However, although the thermal properties
of Green functions in the Rindler, de Sitter and Schwarzschild spaces were
considered \cite{a8}, finite-temperature effective potential and symmetry
breaking in the static spaces with horizons were not investigated.

The present paper studies the quantum theory of a scalar field in the static
de Sitter space-time at arbitrary temperature denoted by $\beta^{-1}$.
The analysis of the scalar case turns out to be rather simple and can help
to understand us the features specific of more realistic gauge theories.

The paper is organized as follows. Section 2 is devoted to the quantization
of a scalar field $\phi$ in the static de Sitter space. The energy operator
in that space can be introduced and divided into two commuting parts,
defined in causally-disconnected regions. This enables
one to formulate the functional integration formalism for the thermal averages
in each region. It turns out that the integration here
goes over the field configurations placed on the compact four-dimensional
space $O_{\beta}$ with Euclidean signature.
This space is the infinitely-sheeted along the "imaginary" time $\tau$
hypersphere $S^4$ of the radius $a$ where
points $(\tau,x^i)$ and $(\tau+\beta,x^i)$ are identified. At the Hawking
temperature, when $\beta= 2\pi a \equiv \beta_H$, the space $O_{\beta}$
becomes a four-sphere $S^4$. In the general case it has conic singularities
where the Killing vector field generating translations along $\tau$ is null.

In Section 3 the finite-temperature effective potential $V(\phi,\beta)$ is
introduced in the framework of the functional integration formalism for
averages.
Studying the spectrum of the Laplace operator on $O_{\beta}$ we are able
to find the expression of the one-loop effective potential as an expansion in
$\beta^{-1}$. We use here zeta-function regularization \cite{a9}, \cite{a9a}.
The suitable forms of $V(\phi,\beta)$ and of the average energy density
$E(\phi,\beta)$
are given for the ground and de Sitter invariant states. It is shown that,
in the limit of asymptotically small space-time curvature, they both coincide
with the vacuum effective potential computed in Minkowski space.

The scaling properties of the theory in the conformally invariant case are
considered in Section 4, where the stress tensor anomaly is obtained
explicitly. Interestingly, it turns out to depend on temperature. At $\beta=
\beta_H$ the standard value of the anomaly is recovered. The possible reasons
of
this circumstance are briefly discussed.

Finally, in Section 5 for a real selfinteracting scalar field
we show the differences of the symmetry breaking
pattern in the ground and de Sitter invariant states.
It is shown that in the ground state a discrete symmetry
of the classical theory is always spontaneously broken, whereas at the
Hawking temperature it can be restored at a certain value of the space
radius $a_{cr}$. Conclusions and remarks are then presented.

Technical details needed for the explicit evaluation of the
zeta-function near
$\beta=\beta_H$ and in the ground state are reported in Appendix A and B,
respectively. The results of Appendix A can be used to estimate the temperature
corrections to the potential near the de Sitter invariant state.

\section{Static de Sitter Space-Time at Nonzero Temperatures}
\setcounter{equation}0

\subsection{Quantization in the static de Sitter space}

De Sitter space-time is a solution of the Einstein equations with a
positive cosmological constant. In the static coordinates
the line element can be written in the form
\begin{eqnarray}
ds^2 & = & \cos^2 \chi dt^2-
a^2(d\chi^2+\sin^2\chi d\theta^2 + \sin^2\chi \sin^2\theta d\xi^2)
\nonumber\\
& \equiv & g_{tt}dt^2-g_{ij}dx^i dx^j~~~(i,j=1,2,3)~~~,
\label{2.1}
\end{eqnarray}
and $-\infty<t<+\infty,~-\pi \leq \chi \leq\pi,~0\leq \theta,\xi \leq \pi$,
$a$ is the radius of space.
The properties of the static coordinates are discussed in \cite{a6}.
One has to mention here that they cover only part of the space-time
and that the regions $|\chi|<\pi /2$ and $|\chi|>\pi /2$ are separated by
the surface ${\cal B}=S^2$ and are causally-disconnected.

We can always choose in de Sitter space a Killing vector field generating
one-parameter group of isometries, a subgroup of $SO(1,4)$. The coordinates
(\ref{2.1}) correspond to the time-like part of a Killing vector field
associated with translations along the time $t$. These coordinates are
restricted by the bifurcate Killing horizon \cite{a16} on which the
Killing vector field is null. It coincides with the event horizons of
observers with trajectories being completely inside the static frame
(\ref{2.1}).
The two-surface ${\cal B}$ is the bifurcation surface that is left unchanged
under the action of the given one-parameter group.

The quantization procedure for a real scalar field in the curved space-time
is given in terms of the commutation relations for the field variables
\cite{a3}
\bee
[\hat{\phi}(x),\hat{\phi}(y)]=0~~~,
\label{2.1a}
\ene
\bee
[\hat{\phi}_{,\mu}(x)d\sigma^{\mu}(x),\hat{\phi}_{,\nu}(y)d\sigma^{\nu}(y)]=0
{}~~~,
\label{2.1b}
\ene
\bee
\int_{\Sigma} f(y)[\hat{\phi}(x),\hat{\phi}_{,\mu}(y)]d\sigma^{\mu}(y)=
if(x)~~~,
\label{2.2}
\ene
where the points $x$ and $y$ belong to a space-like hypersurface $\Sigma$,
such that the Cauchy data on $\Sigma$ define uniquely a solution of the
classical equation in the whole space-time. For the static spaces we
can introduce the energy operator $\hat{H}$ which is associated with
a generator of the unitary transformations of the field $\hat{\phi}$ under
translations along the time coordinate $t$. In the static
de Sitter space (\ref{2.1}) $\hat{H}$ depends on the time
component $\hat{T}_t~^t$ of the energy momentum tensor
\bee
\hat{H}=\int_{t=const}\sqrt{-g}~d^3x \hat{T}_t~^t~~~.
\label{2.3}
\ene
It is splitted into two parts $\hat{H}_1$ and $\hat{H}_2$ depending on the
field variables and acting in the regions $|\chi | \leq \pi /2$
and $|\chi | \geq \pi /2$ respectively ($g$ is the determinant of the metric
(\ref{2.1})). For the model of the real
self-interacting scalar field with the action and the energy momentum tensor
given respectively by
\bee
S=\int d^4x ~ \sqrt{-g}~ \left(\frac{1}{2}\partial_{\mu}\phi\partial^{\mu}\phi-
V(\phi)\right)~~~,
\label{2.4}
\ene
and
\bee
T_{\mu\nu}=2(-g)^{-1/2}
\delta S / \delta g^{\mu\nu}~~~,
\label{2.4b}
\ene
from (\ref{2.1a})-(\ref{2.2}) it follows
\bee
[\hat{H}_1,\hat{H}_2]=\frac{i}{2}\int_{\cal B}
d\sigma^{\alpha}\sqrt{ g_{tt}}(\partial_t\hat{\phi}\partial_{\alpha}
\hat{\phi}+\partial_{\alpha}\hat{\phi}\partial_t\hat{\phi})=0~~~,
\label{2.4a}
\ene
where $d\sigma^{\alpha}$ is the surface element of $\cal B$;
the operators $\hat{H}_1$ and $\hat{H}_2$ commute
because the time component of  the metric tensor $g_{tt}$ vanishes
on the bifurcation surface. In particular, the last equality shows explicitly
that there is no energy
exchange between the two causally-disconnected regions.

\subsection{Functional integration formalism for the averages}

We can choose now (in an oscillator approximation) the creation and
annihilation operators of particles
associated with the Hamiltonian (\ref{2.3}), this allows us to
construct the representation of the commutation relations
(\ref{2.1a})-(\ref{2.2})
given on the corresponding Fock space.\\ Let us consider a canonical ensemble
of such particles at temperature $\beta^{-1}$ in one of the
causally-connected regions, when $|\chi | < \pi/2$, for instance.
The thermally averaged value of a physical variable $\hat{{\cal O}}$
measured in this region reads
\bee
<\hat{{\cal O}}>_{\beta}=Z_{\beta}^{-1}Tr(\hat{{\cal
O}}e^{-\beta\hat{H}_1})~~~,
\label{2.5}
\ene
where $Z_{\beta}$ is the partition function determined by the eigenvalues
$E_n$ of the operator $\hat{H}_1$
\bee
Z_{\beta}=Tr(e^{-\beta\hat{H}_1})=\sum_n e^{-\beta E_n}~~~.
\label{2.6}
\ene
The parameter $\beta^{-1}$ coincides with the local temperature measured
by the observer being at the origin of the static coordinates at $\chi =0$
and the average (\ref{2.5}) does not depend on the behavior of the
system in the rest of space.

To obtain the functional integral representation for the average values
(\ref{2.5}),
let us make the coordinates $x^i$ discrete (with intervals
$\Delta x^i$) on the surface $t=const$. Then, in the causally connected region
$|\chi | <\pi /2$, the transition amplitude from the state $|\phi'>$ to the
state
$|\phi>$ for the infinitesimal {\it imaginary-time} $\epsilon$ turns out
to be
\bee
U_{\epsilon}(\phi,\phi')=<\phi|e^{-\epsilon\hat{H}_1}|\phi'>=
\lim_{\Delta x^i \rightarrow 0}\prod_{x}\left(\frac{\sqrt{-g(x)}g^{tt}(x)
\Delta x^1\Delta x^2\Delta x^3}{2\pi\epsilon}\right)^{1/2}e^{-S_{\epsilon}
(\phi,\phi ')}~~~,
\label{2.7}
\ene
where
\begin{eqnarray}
S_{\epsilon}(\phi,\phi ')=\frac{\epsilon}{2}\sum_{x}\sqrt{-g(x)}
\Delta x^1\Delta x^ 2\Delta x^3 \left[g^{tt}(x)\left(\frac{\phi(x)-\phi'(x)}
{\epsilon}\right)^2 \right.+
\nonumber\\
\left.
+g^{ii}(x)\left(\frac{\phi(x)-\phi'(x+
\Delta x^i))}{\Delta x^i}\right)^2+V(\phi)\right]~~~.
\label{2.8}
\end{eqnarray}
According to this definition the functional $\Psi(\epsilon,\phi)=
\int d\phi' U_{\epsilon}(\phi,\phi')\Psi(\phi')$ has the following properties:
\bee
\Psi(\epsilon,\phi)|_{\epsilon=0}=\Psi(\phi)~~~,
\label{2.9}
\ene
\bee
-\partial_{\epsilon}\Psi(\epsilon,\phi)|_{\epsilon=0}=\hat{H}_1\Psi(\phi)~~~.
\label{2.10}
\ene
$\hat{H}_1$ is the Hamiltonian in the region $|\chi | < \pi /2$
\bee
\hat{H}_1=\int_{|\chi | <\pi /2} \sqrt{-g}~d^3x \left({1 \over 2} \hat{\Pi}^2 +
{1 \over 2} g^{ii}(\partial_{i}\hat{\phi})^2+V(\hat{\phi})\right)~~~,
\label{2.10a}
\ene
connected with the stress tensor via the formula (\ref{2.3}). On the surface
$t=const$ (that is on $S^3$) $\hat{\Pi}=(g_{tt})^{-1/2} \partial _t\hat{\phi}$
is the quantity proportional to the canonical momentum in the
coordinate representation
\bee
\hat{\Pi}(x) ={1 \over i(det g_{ij})^{1/2}}
{\delta \over \delta\phi(x)}~~~.
\label{2.10b}
\ene
The transition amplitude for the final imaginary-time interval $\beta$
is given by the integral
\bee
U_{\beta}(\phi,\phi')=\int D\tilde{\phi} e^{-S_{\beta}(\phi,\phi')}~~~,
\label{2.11}
\ene
where
\bee
D\tilde{\phi}=\lim_{\triangle \tau,\Delta x \rightarrow 0}
\prod_{0\leq\tau_i\leq
\beta}\prod_{x} \left(\frac{\sqrt{-g} g^{tt}\Delta x^1\Delta x^2\Delta x^3}
{2\pi\triangle\tau}\right)^{1/2} d\tilde{\phi}(x,\tau_i)~~~,
\label{2.12}
\ene
\bee
S_{\beta}(\phi,\phi')_{\triangle\tau,\Delta x\rightarrow 0}=
\int_{0\leq\tau\leq \beta} \sqrt{g_{tt}~detg_{ij}}~ d\tau
d^3 x \left[{ 1 \over 2}g^{tt}(\partial_{t}\tilde{\phi})^2
+\frac 12 g^{ii}(\partial_{i}\tilde{\phi})^2 +V(\tilde{\phi})\right],
\label{2.13}
\ene
with the boundary condition $\tilde{\phi}(x,\tau=\beta)=\phi(x)$,
$\tilde{\phi}(x,\tau=0)=\phi'(x)$.
The representation for the average value of an operator $\hat{{\cal O}}$
follows
from (\ref{2.5}) and (\ref{2.6})
\bee
<\hat{{\cal O}}>_{\beta}=Z_{\beta}^{-1}\int D\phi {\cal O}[\phi] e^{-S_{\beta}
(\phi)}~~~,
\label{2.14}
\ene
\bee
Z_{\beta}=\int D\phi e^{-S_{\beta}(\phi)}~~~,
\label{2.14a}
\ene
where $D\phi\equiv d\phi D\tilde{\phi}$ and $S_{\beta}(\phi)\equiv S_{\beta}
(\phi,\phi)$. From the definition (\ref{2.13}),
the integration in (\ref{2.14}) goes over the field variables placed on the
compact space $O_{\beta}$ with line element
\begin{eqnarray}
ds^2 & = & \cos^2 \chi d\tau^2+
a^2(d\chi^2+\sin^2\chi d\theta^2 + \sin^2\chi \sin^2\theta d\xi^2)~~~,
\label{2.15}
\end{eqnarray}
which is the Euclidean form of the line element (\ref{2.1}), and the periodic
parameter $\tau$ ranging from 0 to $\beta$.

When $\beta=\beta_H$ the space $O_{\beta}$ is the four-dimensional
hypersphere $S^4$. The two-point thermal Green function defined in agreement
with (\ref{2.14}) at $\beta=\beta_H$ coincides with the Green function of
the de Sitter-invariant quantum state that also turns out in static
coordinates to be periodic
analitic function of the imaginary time \cite{a5} with period $2\pi a$.
This state
is the vacuum, but its field excitations, which are defined in a de Sitter
invariant way \cite{a3}, cannot be interpreted as particles
of a certain energy.
All observers moving freely register this state as a thermal equilibrium at
the same temperature $\beta_H^{-1}=(2\pi a)^{-1}$ \cite{a5}.
Let us point out that thermal equilibrium at the Hawking temperature only in
the given part of static frame
($|\chi | <\pi /2$) does not mean the de Sitter-invariant
vacuum because the quantum state of the system in the other casually
independent part of space ($|\chi |>\pi /2$) can be quite arbitrary.

If $\beta = n^{-1}\beta_H~~(n=1,2,..)$,
the integration in the representation (\ref{2.14}) for the averages goes over
the fields on the hypersphere $S^4$ on which the points $(\tau,x^i)$ and
$(\tau+\beta,x^i)$ are identified. Such space is
an orbifold \cite{a7}. At zero temperature $O_{\beta}=O_{\infty}$ and is
an infinitely-sheeted sphere $S^4$. For the arbitrary temperatures $O_{\beta}$
is the factor space of $O_{\infty}$ over the ciclic rotation group with
period $\beta$ leaving the two-surface ${\cal B}$ unchanged.
In all the points out of
${\cal B}$ it has
the geometry of an hypersphere but in the domain of $\cal B$, when $|\chi |
\rightarrow \pi /2$, it looks like the product space
$cone\otimes S^2 $. The volume of $O_{\beta}$ is $\beta {\cal V}$ where
${\cal V}$ is the volume of the spatial part of space-time
(${\cal V}=4\pi a^3/3$).

\section{The Effective Potential}
\setcounter{equation}0

\subsection{Basic formalism}

Phase transitions in curved spaces at arbitrary temperatures can be
investigated
as in the flat one applying the effective potential method.
The effective potential $V(\varphi,\beta)$ in our case can be introduced
via the path
integral representation for the partition function (\ref{2.14a}). For this
purpose,
let us consider in (\ref{2.14a}) the "static" part $\varphi \equiv
(\beta {\cal V})^{-1}
\int_{O_{\beta}} \sqrt{g}~d^4 x \phi(x)$ of the field variables on $O_{\beta}$
($g$ is the determinant of the metric (\ref{2.15}))
$$
Z_{\beta}=\int D\phi e^{-S_{\beta}(\phi)}=\int D(\varphi+\phi') e^{-S_{\beta}
(\varphi+\phi')}=$$
\bee
= N \int d\varphi e^{-\beta {\cal V} V(\varphi, \beta)}~~~ ,
\label{3.1}
\ene
where $N$ is a normalization constant. The potential $V$ is defined by
the integral
\bee
e^{-\beta {\cal V} V(\varphi, \beta)}\equiv \int D\phi' e^{-S_{\beta}(\varphi+
\phi')}~~~,
\label{3.2}
\ene
over the fields obeying the condition
\bee
\int_{O_{\beta}} \sqrt{g}~ d^4x \phi'(x)=0~~~ .
\label{3.3}
\ene
If $V(\varphi, \beta)$ is a known function of $\varphi$, the partition function
can be found from (\ref{3.1}) by the method of stationary phase. The points
$\varphi_i$ of minima of $V(\varphi,\beta)$ correspond to various field
configurations with the average field strength in the
considered volume $\cal V$
equal to $\varphi_i$ in the one-loop approximation.
The real part of $V(\varphi,\beta)$ is a sum of the
classical potential energy $V(\varphi)$ and of the
quantum corrections to it. If a
field configuration $\varphi_i$ is unstable, then $V(\varphi_{i},\beta)$
has a nonvanishing imaginary part determining its decay-rate \cite{a12}.

To calculate the one-loop effective potential, one has to expand the
functional $S_{\beta}(\varphi+\phi')$ in (\ref{3.2}) on $\phi'$, taking
into account the condition (\ref{3.3}), and to approximate it by the
expression
\bee
S_{\beta}(\varphi+\phi')=({\cal V} \beta)V(\varphi) + \frac 12 \int_{O_{\beta}}
\sqrt{g}~d^4x~\phi'(x) \hat{Q}(\varphi)\phi'(x)~~~,
\label{3.4}
\ene
where $\hat{Q}(\varphi)\equiv-\Box + V''(\varphi) $ ($\Box$ is the Laplace
operator defined on $O_{\beta}$). The integration in (\ref{3.2}) can be
performed as usual if we use the completeness of the eigenfunctions
$\psi_n(x)$ of $\Box$, so that the field $\phi(x)$ can be expanded as
\bee
\phi(x)=\sum_{n} \phi_n \psi_n(x)~~~,
\label{3.5}
\ene
where the eigenfunctions are normalized as follows
\bee
\int_{O_{\beta}} \sqrt{g}~d^4x~\psi_n(x)\psi_m(x)= \delta_{n,m}~~~,
\label{3.6}
\ene
and change the measure (\ref{2.12}) by the measure $D\phi=\prod_{n}
(2\pi)^{-1/2} \mu
d\phi_n$ with  $\mu$ a normalization constant . Integrating over $\phi_n$
we get from (\ref{3.2})
\bee
V(\varphi,\beta)= V(\varphi)+{1 \over 2\beta {\cal V}}
\left[\log(\det(\mu^{-2}\hat{Q}))-
\log(\mu^{-2}V''(\varphi))\right]~~~.
\label{3.7}
\ene
According to (\ref{3.3}), we eliminate from $V(\varphi,\beta)$ the contribution
of zero mode of the Laplace operator. The last term is important for analitical
properties of the effective potential when the space-time curvature is large.

If all field configurations are stable, the
one-loop partition function can be derived from (\ref{3.1}) considering
the minima with the zero imaginary part $Im~V(\varphi_i,\beta)=0$.
In the given approximation it turns out to be
\bee
Z_{\beta}=\sum_{i}\left({2\pi N^2 \over
V''(\varphi_i)\beta {\cal V}}\right)^{\frac 12}
e^{-\beta {\cal V} V(\varphi_i,\beta)}~~~.
\label{3.8}
\ene
Taking into account the normalization of the zero mode $\varphi$ in (\ref{3.1})
and (\ref{3.5}),(\ref{3.6}) and (\ref{3.7}) we can substitute $N$
with $\mu(\beta {\cal V})^{1/2}$ and represent $Z_{\beta}$ in the form
\bee
Z_{\beta}=\sum_{i} e^{-\beta {\cal V} \tilde{V}(\varphi_i,\beta)}~~~,
\label{3.9}
\ene
where
\bee
\tilde{V}(\varphi,\beta)=V(\varphi)+{1 \over 2\beta {\cal V}}\log\left(
\det(\mu^{-2}\hat{Q})\right)~~~ .
\label{3.10}
\ene
It is obvious that in the flat-space limit, when the radius and volume of
space tend to infinity, both the quantities $V(\varphi,\beta)$ and
$\tilde{V}(\varphi,\beta)$ coincide.

The field average $<\hat{\phi}(x)>_{\beta}$ in the one-loop approximation
can be found from (\ref{2.14}) in a similar way, and is
\bee
<\hat{\phi}(x)>_{\beta}=\sum_{i} P_{i}(\beta)~\varphi_i~~~,
\label{3.11}
\ene
where the coefficients
\bee
P_i(\beta)={e^{-\beta {\cal V} \tilde{V}(\varphi_i,\beta)} \over
\sum_{k}e^{-\beta {\cal V} \tilde{V}(\varphi_k,\beta)} }~~~,~~~~
\sum_{i}P_i(\beta)=1~~~,
\label{3.11a}
\ene
in the equilibrium state are the probabilities for a given field
configuration $\varphi_i$ to appear. From (\ref{3.9}) we can also obtain the
average energy as the sum
\bee
<\hat{H}>_{\beta}=-{\partial \over \partial \beta}\log Z_{\beta}=
\sum_{i}P_i(\beta)~{\cal V} E(\varphi_i,\beta)~~~,
\label{3.12}
\ene
where the quantities
\bee
E(\varphi_i,\beta)={\partial \over \partial \beta}
\left(\beta \tilde{V}(\varphi_i, \beta)\right)~~~,
\label{3.13}
\ene
are the energy densities of the configurations $\varphi_i$.

In the trivial case of a free scalar field the effective potential $V(\varphi,
\beta)$ has only one minimum at $\varphi=0$, as in the classical theory,
because the determinant in (\ref{3.7}) does not depend on the field $\varphi$.

\subsection{Zeta-function}

To regularize the determinant in Eq.(\ref{3.7}), the zeta-function method
\cite{a9}
can be used because the eigenvalues $\lambda_{n,m}$ of the operator
$-\Box+V''(\varphi)$
on $O_\beta$ can be found exactly. They are characterized by two
nonnegative numbers $n$ and $m$ and depend on the temperature $\beta^{-1}$
\bee
\lambda_{n,m}(\beta)=a^{-2}\left(n-m+(\beta_H/\beta)m\right)\left(n-m +
(\beta_H/\beta) m+3\right)+ V''(\varphi)~~~,
\label{3.14}
\ene
$$
n=0,1,2,....;~~~m=0,1,...,n;
$$
The multiplicity $g_{n,m}$ of the eigenvalue $\lambda_{n,m}$ is
$(n-m+1)(n-m+2)$ if $m\neq0$, and $(n+1)(n+2)/2$ for $m=0$. In the case of
the Hawking temperature $\beta_H^{-1}$ this operator turns into the operator
on the hypersphere $S^4$ with $\lambda_n
=a^{-2}n(n+3)+V''(\varphi)$ and the multiplicity
\bee
g_n=\sum_{m=0}^n g_{n,m}={1 \over 6}(n+1)(n+2)(2n+3)~~~.
\label{3.14a}
\ene

The renormalized $\log\det (\hat{Q} \mu^{-2})$, the effective potential
$V(\varphi,\beta)$ and the average energy $E(\varphi,\beta)$
expressed in terms of the generalized zeta-function
\begin{equation}
\zeta(z,\beta) \equiv \sum_{n=0}^{\infty} \sum_{m=0}^{n} g_{n,m} \left(a^2
\lambda_{n,m} \right)^{-z}~~~.
\label{3.15}
\end{equation}
now read
\bee
\log\left(\det(\mu^{-2}\hat{Q})\right)
=-\left[\zeta'(0,\beta)+\log(\mu^2 a^2) \zeta(0,\beta)
\right]~~~,
\label{3.15a}
\ene
\bee
V(\varphi,\beta)=V(\varphi)-{1 \over 2\beta {\cal V}}\left[\zeta'(0,\beta)+
\log(\mu^2a^2)
\zeta(0,\beta)+\log\left(V''(\varphi)\mu^{-2}\right) \right]~~~,
\label{3.16}
\ene
\bee
E(\varphi,\beta)=V(\varphi)-{1 \over 2 {\cal V}}{\partial \over \partial \beta}
\left[\zeta'(0,\beta)+\log(\mu^2 a^2) \zeta(0,\beta) \right]~~~,
\label{3.17}
\ene
where $\zeta'(z,\beta)\equiv {d \over dz}\zeta(z,\beta)$.

Let us find a more suitable form for the zeta-function (\ref{3.15}).
If we express $(a^2\lambda_{n,m})^{-z}$ as an expansion with
parameter $\Delta\equiv 9/4-a^2 V''(\varphi)$ and point out that
$\sum_{n=0}^{\infty} \sum_{m=0}^{n} f(m,n) =
\sum_{n=0}^{\infty} \sum_{m=0}^{\infty} f(m,n+m)$, we can immediately
perform the summation on the $n$ index in (\ref{3.15}) obtaining
\begin{eqnarray}
\zeta(z,\beta) & = & \sum_{k=0}^{\infty}C_k(z){\Delta}^k
\left\{ \sum_{m=0}^{\infty}
\left[\zeta_R\left(2z+2k-2,(\beta_H/\beta) m+3/2\right)\right.\right.
\nonumber\\
&-&2(\beta_H/\beta) m\zeta_R\left(2z+2k-1,(\beta_H/\beta) m+3/2\right)
\nonumber\\
&+&\left.\left( (m \beta_H / \beta)^2-\frac 14 \right)
\zeta_R\left(2z+2k,(\beta_H/\beta)m+3/2\right)\right]
\nonumber\\
&-&\left.
\frac 12 \left[\zeta_R(2z+2k-2,3/2)-\frac 14 \zeta_R(2k+2z,3/2)\right]
\right\}~~~,
\label{3.18}
\end{eqnarray}
where the coefficients $C_k(z)$ are defined by
\bee
C_k(z) \equiv {\Gamma(z+k) \over k!\Gamma(z)}~~~.
\label{3.19}
\ene
With the following integral representation for the Riemannian
$\zeta_{R}$-function
\bee
\zeta_R(s,a)={1 \over \Gamma(s)}\int_{0}^{\infty}
{y^{s-1}e^{-ay} \over 1-e^{-y}} dy~~~,
\label{3.20}
\ene
we are able to sum up over $m$. For instance, one can get
\bee
\sum_{m=0}^{\infty}\zeta_R(s, (\beta_H/\beta) m+3/2)={1 \over \Gamma(s)}
\int_{0}^{\infty}
{y^{s-1} \over 1-e^{-y}}e^{-\frac 32 y} {1 \over 1-e^{- (\beta_H/\beta) y}}
dy~~~,
\label{3.21}
\ene
for $Re~s > 2$. Thus, inserting (\ref{3.21}) in (\ref{3.18}) we have
\begin{eqnarray}
\zeta(z,\beta)=
\sum_{k=0}^{\infty}C_k(z){\Delta}^k\int_{0}^{\infty}dy~{y^{2z+2k-3} \over
1-e^{-y}}e^{-\frac 32 y}\left\{ {\frac 12 \coth\left({y \beta_H\over 2\beta}
\right) \over \Gamma(2z+2k-2)}\right.
\nonumber\\
-\left.  {\left({y \beta_H \over 2 \beta}\right) \sinh^{-2}
\left({y \beta_H \over 2 \beta}\right)
\over \Gamma(2z+2k-1)}+{ \left({y \beta_H \over 2\beta}\right)^2
\coth\left({y \beta_H \over 2 \beta}\right)
\sinh^{-2}\left({y \beta_H \over 2 \beta}
\right)-{y^2\over 8}\coth\left({y\beta_H \over 2\beta}\right)
\over \Gamma(2z+2k)}\right\}~~~.
\label{3.22}
\end{eqnarray}
Note that from (\ref{3.22}), if the variable $z$ is close to zero,
$\zeta(z,\beta)$ is determined by the behavior of the integrand only in the
vicinity of the lower limit of integration. In this case one can use in
(\ref{3.22}) the definition of the Bernoulli numbers $B_n$
\bee
{x \over e^x-1}=\sum_{n=0}^{\infty}{ B_n \over n!} x^n~~~,
\label{3.22a}
\ene
(valid for $|x|<\pi $), to get a representation
for $\zeta(z,\beta)$ as a series of odd powers of the temperature
\begin{eqnarray}
\zeta(z,\beta)& =& {\beta \over \beta_H }\sum_{k=0}^{\infty} C_k(z) \Delta^k
\sum_{n=0}^{\infty} {B_{2n} \over (2n)!} \left({\beta_H \over \beta}\right)
^{2n} \times
\nonumber\\
\times {\Gamma(2z+2k+2n-1) \over \Gamma(2z+2k)} &  &
\left[\zeta_R(2z+2k+2n-3,3/2)
-\frac 14 \zeta_R (2z+2k+2n-1,3/2)\right]~~~.
\nonumber\\
\label{3.23}
\end{eqnarray}
This representation holds for $z$ close to zero and can be applied
to compute $\zeta(0,\beta) \equiv \lim_{z \rightarrow 0} \zeta(z,\beta)$
and its
first derivative, obtaining for $\zeta(0,\beta)$ the exact simple expression
\bee
\zeta(0,\beta)={\beta \over \beta_H} \left[ {\left(51-60 (\beta_H/\beta)^2-
8(\beta_H/\beta)^4 \right) \over 2880} +
{\left( 2(\beta_H/\beta)^2-3 \right) \over 24}\Delta
+{1\over 12}\Delta^2 \right]~~~.
\label{3.24}
\ene
At the Hawking temperature this result coincides with the expression
obtained by other authors \cite{a1}.
To compute the first derivative of (\ref{3.23}) we observe that
\bee
\left.{d \over dz}(z \zeta_R (z+n+1,a)) \right| _{z=0}={ (-1)^{n+1} \over n!}
{ d^n \over da^n}\psi(a)~~~,
\label{3.24a}
\ene
with $n$ integer $>0$. Unfortunately, as far as $\zeta'(0,T)$
is concerned, it can only be expressed in terms of an expansion
$$
\zeta'(0,\beta)  = - {2 \over T} \left[ \zeta_{R}\left(-3,{3 \over 2}\right)
- { 1 \over 4} \zeta_{R}\left(-1,{3 \over 2}\right) \right]
- {2 \over T} \left[ \zeta_{R}'\left(-3,{3 \over 2}\right)
- { 1 \over 4} \zeta_{R}'\left(-1,{3 \over 2}\right) \right]
$$
$$
+ {1 \over T}\left( \Delta + {T^2 \over 6} \right) \left[
\zeta_{R}\left(-1,{3 \over 2}\right) + { 1\over 4} \psi\left( { 3 \over 2}
\right) \right] + { 1 \over T} \left({ \Delta^2 \over 36}+{\Delta \over 4 }+
{\Delta T^2 \over 12}-{ T^4 \over 120} \right)
$$
$$
-{ 1\over T} \sum_{n=0}^{\infty} \left[ {\Delta^2 \over 12} B_{2n}+
\Delta T^2 {B_{2 n+2} \over (2n+2)(2n+1)} + 2 T^4 { B_{2n+4}\over
(2n+4)(2n+3)(2n+2)(2n+1)} \right]\times
$$
$$
\times{T^{2n} \over 2n! }
\left[ (2n+2)(2n+1) \psi^{(2n)}\left({3 \over 2} \right) -  { 1\over 4}
\psi^{(2n+2)}\left({3 \over 2} \right) \right]
$$
$$
- { 2 \over T} \sum_{k=3}^{\infty} \sum_{n=0}^{\infty}
B_{2n} { (\sqrt{\Delta})^{2k} \over 2k! }
{T^{2n} \over 2n!} \left[ (2n+2k-2)(2n+2k-3) \psi^{(2n+2k-4)}\left({3 \over 2}
\right) \right.
$$
\begin{equation}
-\left.{ 1\over 4} \psi^{(2n+2k-2)}\left({3 \over 2} \right) \right],
\label{3.25}
\end{equation}
where $T\equiv\beta_H/\beta$.
The last equations (\ref{3.24}) and (\ref{3.25}), once inserted into
(\ref{3.16}), define explicitly the effective potential as an expansion
in the temperature $\beta^{-1}$.  This expansion is especially useful to
investigate  the potential $V(\varphi,\beta)$ at the low temperatures.
Another expansion of $V(\varphi,\beta)$ around the Hawking temperature can be
found in Appendix A.
However, in the most interesting cases we are going to consider,
the potential can be written in a more suitable integral form.

\subsection{Vanishing temperature and Hawking temperature}

The effective potential for the space of radius $a$ at the Hawking temperature
$\beta_H^{-1}$ can be found
from (\ref{3.16}) substituting in it the expressions of
$\zeta'(0,\beta_H)$ obtained in \cite{a1}, and of $\zeta(0,\beta_H)$ from
(\ref{3.24}). It reads
$$
V(\varphi,\beta_H)=V(\varphi)
-{3 \over (4\pi)^2 a^4}\left[-\frac 13 \left(\int_{\frac 12}^{\frac 12 +
\sqrt{\Delta}}+
\int_{\frac 12}^{\frac 12 - \sqrt{\Delta} }\right)
u(u- \frac 12)(u-1)\psi(u) du \right.
$$
\bee
\left. +  {1 \over 12}\Delta^2 + {1 \over 72} \Delta
+ \log(\mu^2 a^2)\left({\Delta^2 \over 12} - {\Delta \over 24} -
{17 \over 2880} \right) + \log\left(V''(\varphi)\mu^{-2}\right) \right]+ const,
\label{3.26}
\ene
where $\psi(u)$ is
the psi-function. We can also derive the average energy density (\ref{3.13})
in this
state by an expansion of $\zeta(z,\beta)$ in powers of
$(\beta_H-\beta)/\beta$ given in Appendix A
$$
E(\varphi,\beta_H)=V(\varphi)+{3 \over (4\pi)^2 a^4}\left[-\frac 18 \Delta^2 +
{41 \over 144}\Delta - {973 \over 5760} \right.
$$
\bee
\left.+{1 \over 12}\left(\frac 94-\Delta\right)\left(\frac 14-\Delta\right)
\left(\psi(3/2+\sqrt{\Delta})+
+\psi(3/2-\sqrt{\Delta})-\log(\mu^2 a^2)\right)
\right].
\label{3.27}
\ene

 From (\ref{3.16}),(\ref{3.17}) and (\ref{3.23}) the effective
potential at zero temperature coincides with the vacuum energy density.
A connection
between zeta-functions at $\beta =\infty $ and $\beta=\beta_H$,
described in Appendix B, implies
$$
V(\varphi,\infty)=V(\varphi)-{3 \over (4\pi)^2 a^4}\left[
\left(\int_{\frac 12}^{\frac 12 + \sqrt{\Delta}}+
\int_{\frac 12}^{\frac 12 -\sqrt{\Delta} }\right)u
\left(u- \frac 12- \sqrt{\Delta}\right) (u-1)\psi(u) du\right.
$$
\bee
\left.+{1 \over 36}\Delta^2 + {7 \over 24} \Delta
+\log(\mu^2 a^2)\left({\Delta^2 \over 12} - {\Delta \over 8}
+{17 \over 960}\right)\right]+const.
\label{3.28}
\ene
At the points of minima the imaginary part $Im~V(\varphi,\infty)$ gives
the decay probability $\Gamma$ of metastable vacuum configurations calculated
in the quasiclassical approximation $\Gamma=$ \\ $-2 Im~ V(\varphi,\infty)$.
When $\sqrt{\Delta}\geq 3/2$ or if $V''(\varphi)\leq 0$,
the integrand in (\ref{3.28}) has the simple poles due to the psi-function
and integration contour
should be chosen so that $Im~V(\varphi,\infty)\leq 0$.
This can be achieved simply by changing $V''(\varphi)$ with $V''(\varphi)-
i\epsilon/2$ ($\epsilon > 0$), which corresponds to go around the poles in the
lower part of the complex plane.

A similar way to regularize the integral part of $V(\varphi,\beta)$
can be
taken at $\beta=\beta_H$, but here the situation is different. The
vacuum energy (\ref{3.28}) is singular when $V''(\varphi)=0$ where both
$E(\varphi,\beta_H)$
and $V(\varphi,\beta_H)$ are finite. The singularity and imaginary part
coming from the integral in (\ref{3.26}) when $3/2\leq\sqrt{\Delta}<5/2$
are totally cancelled by the last term $\log(V''(\varphi)\mu^{-2})$.
Consequently, in the vacuum state one has instability when $V''(\varphi)\leq
0$;
whereas at the Hawking temperature, when $V''(\varphi)\leq -4a^{-2}$
(or $\sqrt{\Delta}\geq 5/2$).

Asymptotic expressions for $V(\varphi,\infty),~V(\varphi,\beta_H)$ and
$E(\varphi,\beta_H)$
at the large radius $a$ are written in the Appendices. One can thus show
that all
three quantities in the limit $a\rightarrow\infty$ coincide with the vacuum
effective potential in Minkowski space
\bee
V_M(\varphi)=V(\varphi)+{1 \over 64\pi^2} \left(V''(\varphi)\right)^2
\left(\log(V''(\varphi)\mu^{-2})-\frac 32\right).
\label{3.30}
\ene
This property can be easily explained observing that the Hawking
temperature ($(2\pi a)^{-1}$) vanishes in the flat-space limit. On the other
hand, the effective potential calculated
at $\beta=\beta_H$ coincides with the one
in a de Sitter invariant state and can be turned, when
$a\rightarrow \infty$, only into the potential in the Poincare-invariant
vacuum state.

To complete the calculation of the
renormalized $V(\varphi,\beta)$ we have to add to
it finite counterterms and express the parameters through the measured
quantities. It will be done for a particular model in Section 5.

\section{Scaling and the Trace Anomaly}
\setcounter{equation}0

Let us consider the conformally invariant scalar field theory with the
potential $V(\phi)=(R/12)~\phi^2$, where $R$ is the scalar curvature,
$R=12a^{-2}$ for de Sitter space-time. The energy operators $\hat{H}$
of conformally related static metrics $\tilde{g}_{\mu\nu}(x)=
\alpha^2(x)g_{\mu\nu}(x)$ have the same eigenvalues \cite{a11}.
In this case, the scale invariance of the unrenormalized partition function
$Z_{\beta}$ follows immediately from the definitions (\ref{2.12}),(\ref{2.13})
of the measure $D\phi$ and the Euclidean action $S_{\beta}(\phi)$ in
(\ref{2.14a}).

In the conformally invariant scalar field theory, the logarithm of the
renormalized partition function is defined by (\ref{3.9}),(\ref{3.15a})
and reads
\bee
\log Z_{\beta}=\frac{1}{2}
{}~\left[\zeta'(0,\beta)+\log(\mu^2 a^2) \zeta(0,\beta)\right]~~~.
\label{4.1}
\ene
For the constant scale transformations of the metric $\tilde{g}_{\mu\nu}(x)=
\alpha^2g_{\mu\nu}(x)$ we have
$\tilde{\lambda}_{n,m}=\alpha^{-2}\lambda_{n,m}~$,$\tilde{\zeta}(z,\beta)=
\alpha^{2z}\zeta(z,\beta)$ and therefore the following equality for the
partition function, as a function of $g_{\mu\nu}$ and the renormalization
parameter $\mu$, holds
\bee
Z_{\beta}(\alpha^2 g_{\mu\nu},\alpha^{-1}\mu)=Z_{\beta}(g_{\mu\nu},\mu)~~~.
\label{4.2}
\ene

In static space-times the thermally averaged energy momentum tensor
does not depend on time and can be determined by functionally differentiating
the free energy $F(\beta)=-\beta^{-1}\log Z_{\beta}$ \cite{a11}
\bee
T_{\mu\nu}(\beta,x)=-{2 \over \sqrt{-g}}{\delta F(\beta) \over \delta
g^{\mu\nu}(x)}~~~
\label{4.3}
\ene
($x^i$ are three spatial coordinates).
Thus, one can write for the integral of its trace ($T_{\nu}~^{\nu}(\beta)$)
over
the spatial volume ${\cal V}$ the following equation
\bee
\int_{{\cal V}}d^3x ~\sqrt{-g}T_{\sigma}~^{\sigma}(\beta,x)=
{2 \over \beta}\int_{{\cal V}} d^3x~{\delta\log Z_{\beta} \over
\delta g^{\mu\nu}
(x)} g^{\mu\nu}(x)=-\beta^{-1}{\partial \over \partial\alpha}
\log Z_{\beta}(\alpha^2 g_{\mu\nu},\mu)|_{\alpha=1}~~~.
\label{4.4}
\ene
Finally, Eqs.(\ref{4.1}) and (\ref{4.2}) give
\bee
\int_{{\cal V}}d^3x ~\sqrt{-g}T_{\sigma}~^{\sigma}(\beta,x)=
-\beta^{-1}{\partial \over \partial\alpha}
\log Z_{\beta}(g_{\mu\nu},\alpha \mu)|_{\alpha=1}
=-{\beta}^{-1}\zeta(0,\beta)~~~.
\label{4.5}
\ene
Substituting here the derived expression (\ref{3.24}) for $\zeta(0,\beta)$
in the conformal case $(\Delta=1/4)$ we get the trace anomaly at the
temperature
$\beta^{-1}$
\bee
{\cal V}^{-1}\int_{{\cal V}}d^3x ~\sqrt{-g}T_{\sigma}~^{\sigma}(\beta,x)=
{1 \over 960 \pi^2 a^4}\left(3+\left(\beta_H /\beta\right)^4 \right)~~~.
\label{4.6}
\ene
Remarkably, it is a function of $\beta^{-1}$ and leads
at the Hawking temperature $\beta_H^{-1}=(2 \pi a)^{-1}$ to the correct trace
anomaly and energy-momentum tensor of the de Sitter-invariant state
\bee
T_{\mu\nu}(\beta_H,x)=(960 \pi^2 a^4)^{-1} g_{\mu\nu}(x)~~~.
\label{4.7}
\ene

It is ordinary believed that the trace anomaly does not depend on the
quantum state in which is the system \cite{a4} because it is determined
by the ultraviolet divergences and is sensible only to the space-time geometry
and to the possible boundaries.

The general finite-temperature quantum field theory in static space-times
has been investigated in \cite{a11}. It has been shown there that infinities,
renormalization, and the trace anomaly are the same as at zero
temperature. However, the effects of horizons that can be crucial for our
analysis were ignored in that work.

The divergences arising in the case of the static de Sitter space can be
investigated for the thermal two-point Green function. Considered
as a function of the imaginary time ranging from zero to $\beta$, it is
given on the compact space $O_{\beta}$ (see (\ref{2.15})) with the conic
singularities near two-surface $\cal B$, which may
effect its unusual thermal properties at short distances.
Analogous thermal Green functions, corresponding to
the Rindler and Schwarzschild metrics, are defined on the spaces
with the same conical structure near the horizons. This is probably true for
the case of every space-time with the bifurcate Killing vector field.
However a detailed analysis of the dependence of
the anomaly on the thermal state, and of the role
of the horizon is outside the aim of the present work.

We should also mention the calculation of the average energy
according to (\ref{3.12}) in terms of the renormalized function $E(\varphi,
\beta_H)$,
Eq.(\ref{3.27}). In the conformal case we are interested in , it is
simply equal
to ${\cal V}E(0,\beta_H)$, with $\Delta=1/4$ and the average value
of the field $\varphi=0$. The energy thus
obtained does not depend on the scale parameter $\mu$.
There is a discrepancy between it and the quantity
$<\hat{H}>_{\beta}\equiv\int d^3x~\sqrt{-g}~T_t~^t(\beta_H,x)$ defined through
the
anomalous energy momentum tensor (\ref{4.7}). However, whereas the first one,
$E(0,\beta_H)$, is defined up to finite renormalization terms,
the quantity $<\hat{H}>_{\beta}$ is totally anomalous and consequently
is of a pure geometrical character and independent of the renormalization
procedures \cite{a4}.

\section{The Model}
\setcounter{equation}0

We study here, as an example, the model of a real quantum scalar field
with symmetrical potential
\bee
V(\phi)=-\frac 12 \sigma^2\phi^2+{\lambda \over 4} \phi^4~~~,
\label{5.1}
\ene
$(\sigma ^2,\lambda >0)$ and compute the effective
potential in the ground and
de Sitter-invariant quantum states.

The discrete symmetry $\phi\rightarrow
-\phi$ inherent in the classical model (\ref{5.1}) is known to be broken
in the ground state in flat space-time: in this case the zero-field
configurations are unstable. The symmetrical phases
correspond to the configurations with zero field strength and their
relevance at nonzero space-time curvature may be found from the results
derived in Section 3.

 From these results we draw immediately the conclusion that there cannot be
stable symmetrical phases in the ground state at any curvature because
$V''(0)=-\sigma^2<0$ and the effective potential has a non-zero imaginary
part at $\phi=0$.
On the other hand, symmetry can be restored at the
Hawking temperature $\beta_H^{-1}$ at a certain value of $a$ if the
following conditions hold:
\bee
V'(0,\beta_H)=0,~~V''(0,\beta_H)\geq 0,~~~V''(0)>-4a^{-2}.
\label{5.2}
\ene
The first condition is always true for this model as far as
$V(\phi)$ depends only on the square of the field. To investigate the second
one
we have to fix the meaning of the constants $\sigma$ and $\lambda$ in terms
of the measurable quantities, obtained for instance in flat-space.

Following the standard renormalization procedure we can eliminate the scale
parameter $\mu$ from $V(\varphi,\beta_H)$, Eq.(\ref{3.26}), by absorbing it
into the definition of the finite counterterms that should be added to the
effective potential. These counterterms have the same structure as
the initial potential (\ref{5.1}). Thus, the renormalized $V(\varphi,\beta_H)$
turns out to be
$$
V(\varphi,\beta_H)=V(\varphi)
-{3 \over (4\pi)^2 a^4}\left[-\frac 13 \left(\int_{\frac 12}^{\frac 12 +
\sqrt{\Delta}}+\int_{\frac 12}^{\frac 12 - \sqrt{\Delta} }\right)
u(u- \frac 12)(u-1)\psi(u) du \right.
$$
\bee
\left. +  {1 \over 12}\Delta^2 + {1 \over 72} \Delta
+ \log\left(V''(\varphi)a^2\right) \right]+ A \varphi^2 + B \varphi^4 + const,
\label{5.3}
\ene
In the limit of asymptotically small curvature ($a\rightarrow\infty$)
(\ref{5.3}) takes the form
\bee
V_M(\varphi)=V(\varphi)+{1 \over 64\pi^2} \left(V''(\varphi)\right)^2
\left[\log(V''(\varphi)a^2)-\frac 32\right] +A\varphi^2 + B\varphi^4 + const,
\label{5.4}
\ene
and the renormalization conditions for it can be chosen as
\bee
V'_M(\varphi)|_{\varphi^2=\sigma^2/\lambda}=0,~~~~
V''_M(\varphi)|_{\varphi^2=\sigma^2/\lambda}=2\sigma^2\equiv m^2.
\label{5.5}
\ene
They just define the positions of minima of the asymptotically flat
$V(\varphi,\beta_H)$ and the physical mass $m$ of the field
as in the classical theory (\ref{5.1}). Moreover, they fix the values
for the constants $A$ and $B$
\bee
A={\lambda\sigma^2 \over 32\pi^2}\left(3\log(2\sigma^2 a^2)+6\right),~~~
B=-{9\lambda^2 \over 64\pi^2}\log(2\sigma^2 a^2).
\label{5.6}
\ene
The flat-space potential (\ref{5.4}) so obtained recovers the already
known result reported in \cite{a15}
$$
V(\varphi,\beta_H)|_{a\rightarrow\infty}=-\frac 12 \sigma^2\varphi^2 +
{\lambda \over 4}\varphi^4 + {\left(3\lambda\varphi^2-\sigma^2\right)^2
\over 64\pi^2}\log\left({3\lambda\varphi^2-\sigma^2 \over 2\sigma^2} \right)+
$$
\bee
+ {21\lambda\sigma^2 \varphi^2 \over 64 \pi^2} -{27\lambda^2\varphi^4
\over 128\pi^2} + const.
\label{5.7}
\ene

The same renormalization conditions (\ref{5.5}) and constants (\ref{5.6})
can be chosen at zero temperature because $V(\varphi,\infty)$ and
$V(\varphi,\beta_H)$ have the same flat-space limit.

We can now investigate the second derivative $V''(0,\beta_H)$ that
follows from (\ref{5.3}),(\ref{5.6}) and takes particularly simple form at
sufficiently large curvature, when $a^2<<\sigma^{-2}$,
\bee
V''(0,\beta_H)=-\sigma^2 + {\lambda \over 16\pi^2a^2}(1+6\gamma)
+ {6\lambda\sigma^2 \over 32\pi^2}\left(2+\log(2\sigma^2a^2)\right)~~~,
\label{5.8}
\ene
where $\gamma \neq 0,577...$ is the Euler constant. As one can see
$V''(0,\beta_H)$
changes sign and becomes positive at some critical value of the radius
$a=a_{cr}$. It can be found neglecting the last term in (\ref{5.8})
with respect to the second one and reads
\bee
a^2_{cr}={(1+6\gamma) \over 8\pi^2}{\lambda \over m^2}~~~.
\label{5.9}
\ene
The third condition (\ref{5.2}) holds if $m^2a^2_{cr}<8$, which is true
for not very large values of $\lambda$.

As a conclusion, we have shown in this paragraph that, while in the ground
state the symmetry is always spontaneously broken, the
stable symmetrical phases can
appear at the Hawking temperature at some finite values of the space-time
curvature. The nature of the given phase transition can be understood by
considering the global structure of the effective potential with the help
of the expressions (\ref{5.3}),(\ref{5.6}).

\section{Conclusions and remarks}

We have evaluated the finite-temperature effective potential
for a scalar field theory in de Sitter space-time.
The expression found enables one to study the symmetry breaking in two of
the most interesting cases: at low temperature, and at a temperature
close to the Hawking one. The analysis is explicitly performed for the
bare scalar potential reported in (\ref{5.1}) and shows how strongly the
presence of the temperature affects the phase transition of the system.

It is well known that in Minkowski space-time the classical symmetry
of a scalar potential under the discrete transformation
$\phi \rightarrow - \phi$ is spontaneously broken by the quantum effects.
Remarkably, at low temperatures the symmetrical phase  under this
transformation
is unstable for every value of the radius $a$, whereas at the Hawking
temperature, this symmetry can be recovered for some finite value of $a$.

For a generalization of these results to more realistic gauge
theory, one has to find
the eigenvalues and multiplicities of the corresponding wave operators
of the bosonic and fermionic fields  on the compact space $O_{\beta}$, which
appear in the integral representation for thermal averages.

Finally, we also study the stress tensor anomaly for the conformally
invariant case and find that it is a function of the thermal quantum state
of the system. The reason of this interesting fact and the possible role of
the horizon here will be investigated separately.

\begin{center}{{\large \bf Acknowledgements}}\\\end{center}

The authors would like to thank Prof. F.Buccella and Dr. G.Esposito
for enlightening discussions and for reading the manuscript. Furthermore,
one of the authors (D.V.F) would also like to thank the University of Napoli
for
hospitality during his visit and Dr.S.N.Solodukhin for valuable remarks.

\newpage
{\appendix \noindent{\large \bf Appendix A. Zeta-function at $\beta\simeq
\beta_H$} \\
\def\theequation{A.\arabic{equation}}
\setcounter{equation}{0}

To discuss the expression of the effective potential near the Hawking
temperature, it is useful to represent $\zeta(z,\beta)$
as an expansion in powers of $(\beta_H-\beta)/\beta$. In fact from
(\ref{3.14}) and (\ref{3.15}), we can write
$$
\zeta(z,\beta) = \zeta(z,\beta_H)+{1 \over 3}  \sum_{k=0}^{\infty} \Delta^k
{\Gamma (z+k) \over \Gamma (z+1) k!}
\left\{ z \zeta_{R}\left(2z+2k-3,{3 \over 2} \right)  \right.
$$
$$
\left.-{1 \over 4} z \zeta_{R}\left(2z+2k-1,{3 \over 2} \right) \right\}
+\sum_{p=1}^{\infty} \sum_{r=0}^{p} \sum_{k=0}^{\infty} \Delta^{k}  (-1)^{p}
2^{p-r} \left( \begin{array}{c} p \\ r \end{array} \right)
\left({\beta_H-\beta \over \beta} \right)^{p+r}
{\Gamma (z+p+k) \over \Gamma (z+1) p! k!}\times
$$
$$
\times
\left\{\sum_{t=0}^{p+r+3} {(p+r)! \over (p+r+3-t)!}
{(t-1)(t-2) \over t!}   (-1)^{t} z \zeta_{R}\left(2z+2k+t-3,{3 \over 2}
\right)\times  \right.
$$
$$
\left( t 2^{1-t}-\left(1 - 2^{1-t} \right) B_{t} \right)
+z \zeta_{R}\left(2z+2k+p+r-2,{3 \over 2}\right) \left(-{B_{p+r+1} \over
(p+r+1) } \right)
$$
$$
+z \zeta_{R}\left(2z+2k+p+r-1,{3 \over 2}\right)\left(  {2B_{p+r+2} \over
(p+r+2) }\right)
$$
$$
+  z \zeta_{R}\left(2z+2k+p+r,{3 \over 2}\right) \left(- { B_{p+r+3} \over
(p+r+3)} + \frac 14 ~{ B_{p+r+1} \over (p+r+1)}\right)
$$
\begin{equation}
\left.- {{ 1 \over 4}  \over (p+r+1)}
\sum_{t=0}^{p+r+1} \left( \begin{array} {c} p+r+1 \\ t
\end{array} \right) (-1)^{t} z \zeta_{R}\left(2z+2k+t-1,{3 \over 2} \right)
\left( t 2^{1-t} - \left( 1 - 2^{1-t} \right) B_{t} \right) \right\}~~~,
\label{A.1}
\end{equation}
where we have used the well known relation
\begin{equation}
\sum_{m=1}^{n} m^{p} = {1 \over p+1} \left( B_{p+1}(n+1)- B_{p+1} \right)~~~,
\label{A.2}
\end{equation}
where $B_{n}(x)$ $(B_{n})$ are the Bernoulli polynomials (numbers).
It is worth pointing out
that the expression (\ref{3.24}) for $\zeta(0,\beta)$, derived from another
expansion (\ref{3.23}), can be also obtained from (\ref{A.1}).

The zeta-function at $\beta=\beta_H$ was found in \cite{a1} and is given by
the series
\begin{eqnarray}
\left.\zeta(z,\beta)\right|_{\beta=\beta_H=1} & \equiv & \zeta_H(z,\Delta)=
\nonumber\\
=\frac 13 \sum_{k=0}^{\infty}C_k(z)\left( \sqrt{\Delta}\right)^{2k}
& \times &
\left[\zeta(2z+2k-3,3/2)-\frac 14 \zeta(2z+2k-1,3/2)\right]~~~.
\label{A.3}
\end{eqnarray}
Its derivative has the following integral representation
\begin{eqnarray}
\zeta'(0,\Delta)& = &
-\frac 13 \left(\int_{\frac 12}^{\frac 12 + \sqrt{\Delta}}+
\int_{\frac 12}^{\frac 12 - \sqrt{\Delta} }\right)u\left(u- \frac 12\right)
\Bigr(u-1\Bigr)\psi(u) du
\nonumber\\
& + & {1 \over 12}\Delta^2 + {1 \over 72} \Delta
+\frac 23 \left[\zeta_R'(-3,3/2)-\frac 14 \zeta_R'(-1,3/2)\right]~~~.
\label{A.4}
\end{eqnarray}
 From (\ref{A.1}) it is quite easy to obtain the approximate expressions
for $\zeta(0,\beta)$ and $\zeta'(0,\beta)$ for $\beta \approx \beta_H$,
in fact we have
\bee
\zeta(0,\beta)  \approx \frac{\Delta^2}{12} - \frac{\Delta}{24}
-\frac{17}{2880} +(1-\beta_H/\beta)\left(\frac{\Delta^2}{12} - \frac{5 \Delta}
{24}+\frac{3}{64}\right)~~~,
\label{A.5}
\ene
\begin{eqnarray}
\zeta'(0,\beta)  \approx \left\{ - {1 \over 3} \left(
\int_{{1\over 2}}^{{1\over 2}+\sqrt{\Delta}} +
\int_{{1\over 2}}^{{1\over 2}-\sqrt{\Delta}} \right)
u (u-{1 \over 2})(u-1)\psi(u) du \right.
\nonumber\\
+\left. \frac{\Delta^2}{12}+\frac{\Delta}{72} +{2 \over 3} \left[
\zeta_{R}'(-3,{3 \over 2}) - {1 \over 4} \zeta_{R}'(-1,{3 \over 2})
 \right] \right\}
+ (\beta_H/\beta-1)\left[\frac {41}{144} \Delta - \frac{{\Delta}^2}{8} -
\frac{973}{5760} \right.
\nonumber\\
+ \left.\frac {1}{192} \left( 16{\Delta}^2-40\Delta +9\right)
\left(\psi(\frac 32+\sqrt {\Delta}) +\psi(\frac 32 -\sqrt {\Delta})\right)
\right]~~~.
\label{A.6}
\end{eqnarray}
Inserting (\ref{A.5}) and (\ref{A.6}) in (\ref{3.16}),(\ref{3.17})
we obtain the expressions for the one-loop effective potential and energy
density at a temperature approaching the Hawking value. The next temperature
corrections can be also estimated.

The asymptotic behavior of $V(\varphi,\beta_H)$ and $E(\varphi,\beta_H)$ at
large $a$ when $-\Delta\approx a^2 V''(\varphi) \gg 1$ can be found from
(\ref{3.16}) and (\ref{3.17}) by the asymptotic form
of the psi-function \cite{a10}.
For instance,
\bee
Re~\left(\psi(1/2+iu)\right)|_{u\rightarrow \infty} = \log u - {1 \over 24}
u^{-2} - {7 \over 960}u^{-4} + O(u^{-8}).
\label{A.7}
\ene
One can thus obtain
$$
V(\varphi,\beta_H)=V(\varphi)+{3 \over (4\pi)^2 a^4} \left[\log\left(|\Delta|
(a\mu)^{-2}
\right)\left({\Delta^2 \over 12}-{\Delta \over 24} -{17 \over 2880}\right)
\right.$$
\bee
\left.-\log \left(V''(\varphi)\mu^{-2}\right)
-{\Delta^2 \over 8}+{\Delta \over 24}\right].
\label{A.8}
\ene

\vskip 0.5cm

\newpage

{\appendix \noindent{\large \bf Appendix B. Zeta-function in the ground
state} \\
\def\theequation{B.\arabic{equation}}
\setcounter{equation}{0}

The expression for the effective potential in the ground state follows
from (\ref{3.16})
\bee
V(\varphi,\infty)=V(\varphi)-{3 \over (4\pi)^2a^4}\left[f'(0,\Delta)+
\log(\mu^2)f(0,\Delta)\right]~~~,
\label{B.1}
\ene
where $f(z,\Delta)\equiv\lim_{\beta\rightarrow \infty}\left( \beta^{-1}
\zeta(z,\beta)\right)$.
Using (\ref{3.23}) one can see that
\bee
f(z,\Delta)=\sum_{k=0}^{\infty}C_k(z){\left(\sqrt{\Delta}\right)^{2k}
\over 2z+2k-1} \left[\zeta(2z+2k-3,3/2)-\frac 14 \zeta(2z+2k-1,3/2)\right]~~~,
\label{B.2}
\ene
since $f(0,\Delta)$ can be found using (\ref{3.24})
we only need to compute the derivative ${d \over dz}
f(z=0,\Delta)$. Remarkably, $f(z,\Delta)$ turns out to be connected
with the zeta-function $\zeta_H(z,\Delta)$ in the de Sitter invariant state.
Comparing (\ref{B.2}) and (\ref{A.3}) one can see that
\bee
{d \over d\sqrt{\Delta}}\left[f(z,\Delta)
\left(\sqrt{\Delta}\right)^{2z-1}\right]=3 \Delta^{z-1}\zeta_H(z,\Delta)~~~,
\label{B.3}
\ene
and consequently
\bee
{d \over d \sqrt{\Delta} }\left(\frac {1}{\sqrt{\Delta}}
f'(0,\Delta)\right)=
{1 \over \Delta}\left(3\zeta_H'(0,\Delta)-2f(0,\Delta)\right)~~~.
\label{B.4}
\ene
This equation has the following solution
\begin{eqnarray}
f'(0,\Delta)& = & \sqrt{\Delta} \int_{0}^{\sqrt{\Delta}}
{dy \over y^2}\left[3\left(\zeta'_H(0,y^2)-\zeta'_H(0,0)
\right)-2\left(f(0,y^2)-f(0,0)\right)\right]
\nonumber\\
&-& \left(3\zeta_H'(0,0)-2f(0,0)\right)~~~.
\label{B.5}
\end{eqnarray}
To find $f(0,y^2)$ we take into account (\ref{3.24}) and (\ref{A.5}), so
obtaining
\begin{eqnarray}
f'(0,\Delta)=\left(\int_{\frac 12}^{\frac 12 + y}+
\int_{\frac 12}^{\frac 12 - y}\right)u\left(u- \frac 12- \sqrt{\Delta}\right)
(u-1)\psi(u) du
\nonumber\\
+{1 \over 36}\Delta^2 + {7 \over 24} \Delta
+ {17 \over 480}-2\left[\zeta_R'(-3,3/2)
- \frac{1}{4}\zeta_R'(-1,3/2)\right]~~~,
\label{B.6}
\end{eqnarray}
where in (\ref{B.6}) we have integrated by parts to eliminate one
integration. This result can be inserted into (\ref{B.1})  and $V(\varphi
,\infty)$ takes the form (\ref{3.28}). The asymptotic form of $V(\varphi
,\infty)$ can be obtained using (\ref{A.7}). It is
\bee
V(\varphi,\infty)=V(\varphi)+{3 \over (4\pi)^2 a^4} \left[\log\left(|\Delta|
(a\mu)^{-2}
\right)\left({\Delta^2 \over 12}-{\Delta \over 8} -{17 \over 960}\right)
-{\Delta^2 \over 8}+{\Delta \over 8}\right],
\label{B.7}
\ene
where $-\Delta\approx a^2 V''(\varphi) \gg 1$.

\end{document}